\documentclass[prc,aps,amsmath,reprint, superscriptaddress, amssymb, nofootinbib]{revtex4-1}
\usepackage{graphicx}
\usepackage[utf8]{inputenc}
\usepackage[T1]{fontenc}
\usepackage{lmodern}
\usepackage{dcolumn}
\usepackage{bm}
\usepackage{amsmath}

\begin{document}
\title{
Initial energy-momentum to final flow: a general framework for heavy-ion collisions
}
\author{Jefferson Sousa}
\affiliation{
Universidade de São Paulo, Instituto de Física,
Rua do Matão 1371,
05508-090
São Paulo, SP, Brazil
}
\author{Jorge Noronha}
\affiliation{%
  Department of Physics, University of Illinois at Urbana-Champaign,
  1110 West Green Street, Urbana, USA, Zip Code: 61801-3003
}
\author{Matthew Luzum} 
\affiliation{
Universidade de São Paulo, Instituto de Física,
Rua do Matão 1371,
05508-090
São Paulo, SP, Brazil
}



	\begin{abstract}
	
The evolution of a relativistic heavy-ion collision is typically understood as a process that transmutes the initial geometry of the system into the final momentum distribution of observed hadrons, which can be described via a cumulant expansion of the initial distribution of energy density and is represented at leading order as the well-known eccentricity scaling of anisotropic flow. We extend this framework to include the contribution from initial momentum-space properties, as encoded in other components of the energy-momentum tensor. We confirm the validity of the framework in state-of-the-art hydrodynamic simulations
 of large and small systems. 
 With this new framework, it is possible to separate the effects of early-time dynamics from those of final-state evolution, even in the case when the distribution of energy does not fully determine subsequent evolution, as for example, in small systems. Specifically, we answer the question of when and how azimuthal correlations from the initial state survive to the final state.
In very small systems such as $p$-$p$, for example, initial momentum degrees of freedom dominate over energy. Thus, even if the system forms a quark-gluon plasma that is well described by hydrodynamics, the usual hydrodynamic picture of the transmutation of initial geometry to final momentum anisotropy is broken.  Nevertheless, we show that the hydrodynamic response to the full energy-momentum tensor can be well understood in a similar manner as larger systems. Additionally, this framework elucidates the generic features of the system's evolution that are responsible for the impressive success of hydrodynamic simulations, but which may still hold even in cases where hydrodynamics is not applicable.

	\end{abstract}

\maketitle

\section{Introduction}

Relativistic heavy-ion collisions probe a diverse range of exotic physical phenomena.  While nominally governed by the fundamental theories of the standard model of particle physics, and especially quantum chromodynamics (QCD), it is usually exceptionally difficult to make direct first-principles computations to describe these phenomena.  Instead, experimental measurements can be used to inform our general understanding of the underlying physics.  That is, theoretical descriptions of these complex dynamical processes inevitably involve models or effective theories, with uncertain aspects and parameters that are constrained by measured observables,  thus giving valuable insight into underlying physics.   Due to the complexity of the collision system, these observables may be simultaneously sensitive to a large number of these unknown  parameters, and it can be difficult to isolate and constrain them individually.  Because of this, it is of considerable utility to characterize the dependence of observed quantities on particular aspects of the underlying processes.  In this way, such effects can be separated and studied.  

A notable example is the framework that directly relates geometric properties of the collision system at early times to the anisotropic momentum distribution of particles at the end of the system evolution, expressed in relations such as the famous eccentricity scaling of elliptic flow
\begin{align}
V_2 &= \kappa_2\varepsilon_2.
\end{align}
Here $V_2$ is the elliptic flow --- a particular (vector) measure of the anisotropy of the final distribution of particles in a collision event.  To the extent that this approximate relation holds, it is proportional to the eccentricity --- a particular (vector) measure of the initial spatial anisotropy $\varepsilon_2$, which contains all relevant information from the early-time state of the system that determines the final elliptic flow.   All relevant information about the subsequent evolution of the system is then encoded in the (scalar) response coefficient $\kappa_2$.  Such a relation is now understood as the leading order in a systematically-improvable expansion framework, with subleading contributions that can clearly be seen in simulations \cite{Gardim:2011xv, Hippert:2020kde}. Once established, these types of relations can be very powerful.  Their validity alone gives non-trivial information about the behavior of the system.  Further, this clean separation of effects makes it possible to extract targeted information from experimental data~\cite{Bhalerao:2011yg}.

Until now, this framework only contains information to connect the initial geometry (via the initial distribution of energy in the system) to the final particle distribution.  Here we propose to extend and generalize the framework to include the effects of other degrees of freedom in the early-time system, which are also expected to contribute to final observables --- namely other components of the energy momentum tensor $T^{\mu\nu}$.  While these additional components are typically believed to have subdominant contribution to measured observables in typical heavy-ion collisions, it is important to characterize their effects for several reasons:  
\begin{enumerate}
	\item 
	Since the initial stages of a collision are not well understood, it is not actually known for certain how large or important these contributions are. It thus becomes interesting to quantify their effects, in order to put limits on their possible value.
	\item
	Even if the effects are typically small, they may have outsized effect in certain situations --- e.g., specially-chosen observables or  observables in certain systems such as collisions between smaller ions.
	\item
	This may give additional insight into the framework itself and the minimal requirements for its validity.  While it is inspired by (and usually tested with) hydrodynamic models, it may in fact be more general, similar to how hydrodynamics itself may be valid even far from equilibrium~\cite{Heller:2015dha,Romatschke:2017vte,Strickland:2017kux}.  Thus, when we see signatures of geometric scaling --- or its generalization here including momentum degrees of freedom --- we can better interpret the implications.
\end{enumerate}

In Section \ref{sec:energy}, we review this framework in its current form and its derivation, being careful to specify the minimal assumptions required.  In Section \ref{sec:ansatz}, we present an ansatz for the inclusion of additional components of $T^{\mu\nu}$; namely, the momentum density and stress tensor.  We provide some additional motivation for this ansatz in Section \ref{sec:thought}.   In Section \ref{sec:validation} we test this proposal using state-of-the-art hydrodynamic simulations.  Finally, we summarize our conclusions and our outlook for the future. 

\emph{Notation}: We use natural units where $\hbar=c=1$. 

%
%
%

\section{Review: Cumulant expansion of initial energy density}
\label{sec:energy}
The starting point is the assumption that knowledge of the energy-momentum  tensor at some time $\tau_0$ is sufficient to predict a particular observable of interest to some desired accuracy.
Specifically, we posit that 
\begin{enumerate}
	\item 
the final momentum-space distribution function $f$ of particles in a collision event is a deterministic functional of the energy momentum tensor $T^{\mu\nu}$  and relevant conserved currents $\{j_i^\mu\}$ at some time $\tau_0$,
\end{enumerate}
\begin{align}
\label{eq:ansatz1}
f(p^\mu, \tau\to\infty) &\equiv E \frac{dN}{d^3p}  = \mathcal F [T^{\mu\nu}(\vec x, \tau_0),\{j_i^\mu(\vec x, \tau_0)\}].
\end{align}
It is typical to use a quantity such as rapidity $y$ or pseudorapidity $\eta$ to characterize the momentum along the direction of the beam, while the transverse momentum can be characterized with a magnitude $p_T$ and azimuthal angle $\phi_p$.  The dependence on angle can be nicely characterized by a Fourier series
\begin{align}
E \frac{dN}{d^3p} = N(\eta, p_T) \sum_{n=-\infty}^\infty V_n(\eta, p_T) e^{-in\phi_p},\label{3}
\end{align}
where the anisotropic flow coefficients $V_n$ are 2D vectors, written here as a number in the complex plane.  Each rotational mode then has its own relation
\begin{align}
V_n &= \mathcal F_n [T^{\mu\nu}(\vec x, \tau_0),\{j_i^\mu(\vec x, \tau_0)\}].
\end{align}

More specifically, we assume that this is a good approximation for describing a particular measurement, which may be true even if the assumption does not hold more generally.

Note that, while this assumption is motivated by the success of hydrodynamic simulations, the validity of hydrodynamics is not necessarily required.  Nor does the validity of hydrodynamics automatically mean that it is a good assumption.  For example any hydrodynamic fluctuations that might be present are neglected.    Likewise, it is useful only to the extent that knowledge of some single-body distribution function $f(p^\mu)$ is useful.   Either multiparticle correlations can be neglected, or are specified with their own analogous relation.

The idea, then,  is to characterize this ``system response'' $\mathcal F$ as precisely as possible.  To do this, it is useful to first characterize the initial condition, in a way that important information can be separated from unimportant information.

The general principle that we will use for this separation invokes the presence of a hierarchy of length scales.  We posit that 
\begin{enumerate}
	\setcounter{enumi}{1}
	\item 
	\label{ansatz2}
the structure of the initial conditions at small scales has less importance for the determination of final observables than structure at larger scales.  
\end{enumerate}
As with the first assumption, Eq.~\eqref{eq:ansatz1}, this is motivated by hydrodynamics simulations --- traditionally hydrodynamics is thought of as a description of long-wavelength modes in a system, such that short-wavelength modes are not relevant.  Again, this does not necessarily mean that the validity of the framework is restricted to hydrodynamic systems, nor is its applicability guaranteed by the validity of hydrodynamics.  Nevertheless, its success in describing simulations has already proven its utility and suggests that the assumptions are justified, at least in typical applications.

This separation into a hierarchy of scales is naturally achieved with a spatial Fourier transform of the initial-state fields.
To make this explicit we first make various approximations, which can later be relaxed.  To start, we assume that additional conserved currents (such as the baryon current) can be neglected.   This is typically a good approximation at the highest collision energies, where chemical potentials are close to zero in a large space-time region of the collision system.  We leave the study of conserved currents to future work.

Second, we neglect the longitudinal dependence of the initial state.  For information on this approximation and how to relax it, see Refs.~\cite{Franco:2019ihq, Li:2019eni}, whose methods can be similarly applied to the results of this work.

It is usually assumed that the most important component of the energy-momentum tensor is the energy density $T^{\tau\tau}$.  The goal of this work is to relax this assumption, which will be done in the following section.   For now, we finally have
\begin{align}
	V_n
 = \mathcal F \left[T^{\tau\tau}(\vec x_\perp, \tau_0)\right],
 \label{eq5}
\end{align}
and we need to characterize only a single scalar field $\rho$, which is a function of 2 spatial dimensions
\begin{align}
\label{eq:rho0}
\rho(\vec x_\perp) = T^{\tau\tau}(\vec x_\perp).
\end{align}

We take a 2 dimensional Fourier transform to obtain a cumulant\footnote{The language of cumulants is borrowed from probability theory by analogy --- if $\rho$ were a probability density, then $W_{n,m}$ would be a cumulant in a more traditional sense.} generating function $W(\vec k_\perp)$
\begin{align}
	\label{generating}
e^{W(\vec k_\perp)} \equiv \int d^2 x_\perp\ \rho(\vec x_\perp)\ e^{-i \vec x_\perp \cdot \vec k_\perp}.
\end{align}
In this way, the behavior of the generating function at small $k = |\vec k_\perp|$ represents properties of the initial condition at large length scales, and vice versa.   So we can naturally create a hierarchical set of quantities as coefficients of a Taylor series, expanded around $k=0$
\begin{align}
W(\vec k_\perp) &= 
\sum_{n,m} \frac 1 {m!} W_{n,m} k^m e^{-in\phi_k},
\end{align}
where we simultaneously decompose the coefficients into rotational modes via Fourier series in angle $\phi_k$. (Recall that $k$ is the magnitude of the Fourier variable $\vec k_\perp$, while $\phi_k$ is its azimuthal orientation).

Therefore, the initial density is fully characterized by the discrete set of cumulants $W_{n,m}$, which are cleanly ordered in terms of the length scales they represent. In fact, cumulants with smaller $m$ represents larger scales, and are more important for determining the final $V_n$ according to ansatz \ref{ansatz2}.   

In addition, they are separated into rotational modes labeled by $n$, which will aid in constructing estimators for $V_n$, which also have specific rotation properties.  Specifically, if the system is rotated by some azimuthal angle, $\phi \to \phi+\delta $, then 
\begin{align}
	V_n &\to V_n e^{in\delta},\nonumber\\
	W_{n,m}&\to W_{n,m}e^{in\delta}
	\label{eq:rotate}
\end{align}

We note also that the cumulants constructed this way are translation invariant,\footnote{The exception is $W_{1,1} = \langle x + iy\rangle$ which represents the energy-weighted center of the system, and contains all existing information about  absolute position.} like the momentum-space observables  $V_n$.  These properties will make it simpler to construct estimators with the correct symmetries.

So finally we can formally write the system response as a function of  cumulants (rather than a functional of the initial density), with the sensitivity of the response ordered by index $m$
\begin{align}
	E \frac{dN}{d^3p}  &=f \left(\{W_{n,m}\}\right),\\
	\frac{\partial f}{\partial W_{n,m}} &\gg \frac{\partial f}{\partial W_{n',m'}}  \quad \text{ for } m < m'.
\end{align}
Note that this is not the same as assuming that the initial density has the property $W_{n,m} \gg W_{n',m'}$.  It is instead a statement about the \textit{system response} to the initial conditions.

In general each cumulant is a dimensionful quantity, which must be compared to some scale in order to construct  estimators for the (dimensionless) flow coefficients $V_n$.  In principle a collision system can have a number of relevant scales.  However, for a given collision system (and especially in a fixed centrality interval) many of the scales do not vary greatly.  So in practice it suffices to make the simple and natural choice of the transverse size $\mathcal R$ of the system as defined by the lowest cumulant
%
\begin{align}
	\mathcal R^2 = W_{0,2} = \left\langle \bigl|\vec x_\perp - \langle \vec x_\perp \rangle_E \bigr|^2\right\rangle_E,
\end{align}
with the brackets representing an energy-weighted average
\begin{align}
	\label{eq:bracket}
	\langle\ldots\rangle_E \equiv \frac{\int d^2 x_\perp \ldots  T^{\tau\tau}(\vec x_\perp)}{\int d^2 x_\perp  T^{\tau\tau}(\vec x_\perp)}.
\end{align}
We can then define dimensionless quantities for the anisotropic ($n\neq0$) cumulants
\begin{align}
	\label{eq:eccentricities}
	\varepsilon_{n,m} \equiv - \frac {W_{n,m}}{\mathcal R^m}.
\end{align}
Finally, we posit that
\begin{enumerate}
	\setcounter{enumi}{2}
	\item 
	\label{ansatz3}
	the system response $f \left(\{\varepsilon_{n,m}\}\right)$ can be expressed as a power series in the anisotropy coefficients $\varepsilon_{n,m}$.
\end{enumerate}

Thus, in the end, we have a systematic double expansion, with terms ordered in importance according to the power series\footnote{The convergence properties of this series are not known. We assume that it is, at least, an asymptotic series.}, as well as the value of $m$ in each factor $\varepsilon_{n,m}$.  One can not only determine a leading order estimator for each harmonic $V_n$, but also systematically improve it with higher order corrections.

The most familiar estimator is for $V_2$.   The lowest cumulant with $n\neq 0$ has $m=2$,\footnote{See Appendix \ref{app:cumulants} for more details about the cumulant expansion.} and so the leading order estimator is a linear relation with a single power of the lowest cumulant with $n=2$:
\begin{align}
\mathcal V^{(est)}_2 &= \kappa_{2,2}\, \varepsilon_{2,2}\\
&= -\kappa_{2} \frac{\left\langle \left(r e^{i\phi} - \langle r e^{i\phi} \rangle\right)^2 \right\rangle_E}{\left\langle \left|  r e^{i\phi} - \langle r e^{i\phi} \rangle\right|^2 \right\rangle_E}\\
&\xrightarrow[]{\text{centered}} -\kappa_{2} \frac{\left\langle r^2 e^{2i\phi}\right\rangle_E}{\left\langle r^2 \right\rangle_E},
\end{align}
where the last line is written in a centered coordinate system with $\langle x\rangle_E = \langle y \rangle_E = 0$, and an unimportant numerical factor has been absorbed into the definition of $\kappa_2$.  To leading order, all relevant information about the initial state is contained in $\varepsilon_{2,2}$, while all relevant information about the system response to this initial state is contained in the response coefficient $\kappa$.

The next correction is either the next order linear term $\varepsilon_{2,4}$, or nonlinear terms with $m<4$ such as $\varepsilon_{1,3}^2$ ($m=3$) or even $\varepsilon_{2,2}|\varepsilon_{2,2}|^2$ (which involves only $m=2$ cumulants but is order 3 in the power series).  The general principles of the cumulant expansion do not dictate which type of correction is more important, but must be verified for  the  system in question.

For $V_3$, the expansion is similar.  There is no possible term involving only $m=2$ cumulants (note that to have the correct rotation property, the sum of $n$ values must add to $3$, so we need at least one odd cumulant in each term\footnote{Recall that $W_{1,1}$ represents the center of  the system, and is not an appropriate estimator for $V_n$.  The lowest translation-invariant cumulant with $n=1$ is $W_{1,3}$.} ).  So the leading estimator is again linear
\begin{align}
	\mathcal V^{(est)}_3 &= \kappa_{3,3}\, \varepsilon_{3,3}\\
	&= -\kappa_{3} \frac{\left\langle \left(r e^{i\phi} - \langle r e^{i\phi} \rangle\right)^3 \right\rangle_E}{\left\langle \left|  r e^{i\phi} - \langle r e^{i\phi} \rangle\right|^3 \right\rangle_E}\\
	&\xrightarrow[]{\text{centered}} -\kappa_{3} \frac{\left\langle r^3 e^{3i\phi}\right\rangle_E}{\left\langle r^2 \right\rangle_E^{3/2}},
\end{align}
where a centered coordinate system is again defined by $\langle r^{i\phi}\rangle_E = 0$.
Corrections again include possible linear ($\varepsilon_{3,5}$) and nonlinear (e.g., $\varepsilon_{2,2}\varepsilon_{1,3}$) contributions.

Other predictors can be more complicated.   For $V_4$, for example, one can have a linear estimator $\varepsilon_{4,4}$, which is order $m=4$, or a quadratic estimator $\varepsilon_{2,2}^2$.   Each possible term is lower in one part of the double expansion and higher in the other, and it is not obvious which is more important in a particular system.  In general, both contributions can be important \cite{Gardim:2011xv}.  

As a summary, by making three independent assertions (labeled 1--3 above), we were able to construct a systematic expansion for estimating the flow coefficients $V_n$, representing the rotational modes of the final particle spectrum.   It is useful to note a few important properties of the resulting estimators:
\begin{itemize}
	\item The symmetries are manifest
	\begin{itemize}
		\item 
		Translation invariance --- the flow coefficients $V_n$ are translation invariant, as are the building block cumulants $W_{n,m}$,
		\item
		Rotations ---  each cumulant, as well as each product of cumulants, have a well defined rotation property (see Eq.~\eqref{eq:rotate}) that can be matched with the relevant flow harmonic.
	\end{itemize}
  \item
  The terms are ordered in importance (according to hypotheses 2 and 3), so that the estimator can be systematically improved to arbitrary order.
\end{itemize}

\section{Ansatz for Including additional components of $T^{\mu\nu}$} \label{sec:ansatz}
In addition to the initial geometric distribution of energy, the final-state momentum distribution of particles can also depend on momentum degrees of freedom in the initial state.  Hydrodynamic evolution, for example, depends on the entire  energy-momentum tensor as initial conditions for the equations of motion.  We therefore would like to relax the assumption made in Eq.\ \eqref{eq5} of Sec.~\ref{sec:energy}, and include other components of the energy momentum tensor $T^{\mu\nu}$.

Our proposal is to include the additional effects at the level of the scalar field $\rho$ of Eq.~\eqref{eq:rho0}.  That is, we write
\begin{align}
	\rho(\vec x_\perp)=T^{\tau\tau}(\vec x_\perp)-\alpha\partial_iT^{\tau i}(\vec x_\perp)+\beta\partial_i\partial_jT^{ij}(\vec x_\perp), \label{eq:ansatz}
\end{align}
and construct a cumulant expansion exactly as before, following Eqs. \eqref{generating} through \eqref{eq:eccentricities} without alteration, and constructing estimators as an ordered power series in the generalized eccentricities $\varepsilon_{n,m}$

The new contributions each come with an associated (dimensionful) response coefficient that encodes information about the system response to these aspects of the initial conditions --- $\alpha$ gauges the importance of momentum density relative to the energy density while $\beta$ represents the relative importance of initial transverse stress.  As with the coefficients $\kappa$ that multiply each term in a given estimator, these new coefficients should depend only on the subsequent evolution of the system rather than any aspect of the initial state.   

To illustrate the results of this we list the lowest-order estimators for elliptic and triangular flow.   We first define some  notation
\begin{align}
	U & \equiv T^{\tau x} + i T^{\tau y},\\
	C &\equiv T^{xx} - T^{yy} + 2 i T^{xy},\\
	\langle \ldots \rangle_U &= \frac{\int d^2 x_\perp \ldots U(\vec x_\perp)}{\int d^2 x_\perp  T^{\tau\tau}(\vec x_\perp)},\\
	\langle \ldots \rangle_C &= \frac{\int d^2 x_\perp \ldots C(\vec x_\perp)}{\int d^2 x_\perp  T^{\tau\tau}(\vec x_\perp)},
\end{align}
so that $U$ is a complex representation of  momentum density while $C$ is a complex representation of (the 2 degrees of freedom of) the traceless part of the 2D stress tensor.  Note that, unlike the brackets of Eq.~\eqref{eq:bracket}, these subscripted brackets do not represent weighted averages, but just convenient ratios that show up in the final cumulants.

The relevant cumulants are then,
\begin{align}
	W_{2,2} \propto 
 &
	\langle r^2e^{i2\phi}
		\rangle_E- 2\alpha\langle re^{i\phi}\rangle_U - 2\beta\langle 1\rangle_C  \nonumber\\ &  -(\langle re^{i\phi}\rangle_E - \alpha\langle 1\rangle_U)^2
\end{align}
\begin{align}
	W_{3,3} \propto 
	&\langle r^3e^{i3\phi}\rangle_E - 3\alpha\langle r^2e^{i2\phi}\rangle_U - 6\beta\langle re^{i\phi}\rangle_C\nonumber\\
	&-\Bigr(\langle re^{i\phi}\rangle_E - \alpha\langle 1\rangle_U\Bigr) \nonumber \\
	&\cdot\Bigr(3[\langle r^2e^{i2\phi}\rangle_E - 2\alpha\langle re^{i\phi}\rangle_U - 2\beta\langle 1\rangle_C ]\nonumber\\ &-2[\langle re^{i\phi}\rangle_E - \alpha\langle 1\rangle_U]\Bigr),
\end{align}
%
and the final estimators are constructed as in the previous section, from Eq.~\eqref{eq:bracket}.  A more detailed derivation of cumulants, including numerical factors is shown in Appendix \ref{app:cumulants}.

As before, we can always simplify these expressions with a judicious choice of coordinate center.  For example, one can choose a coordinate system where $W_{1,1} = \langle r e^{i\phi}\rangle - \alpha \langle 1 \rangle_{U} = 0$.   Note that if the net momentum is not zero, then $\langle 1\rangle_U \neq 0$, and this is not the same center of coordinates where $\langle re^{i\phi}\rangle = 0$ that simplifies the expression for the denominator $\mathcal R$.   This coordinate center is then also dependent on response coefficient $\alpha$.  Typically, however, the net transverse momentum is negligible, so the two choices of coordinate center coincide.  Nevertheless, the recentered estimators can then be written as
\begin{align}
\label{eq:estimators}
	\mathcal V^{(est)}_2 
	&= -\kappa_2 \frac
	{\langle r^2e^{i2\phi}
		\rangle_E- 2\alpha\langle re^{i\phi}\rangle_U - 2\beta\langle 1\rangle_C}
	{\left\langle \left|  r e^{i\phi} - \langle r e^{i\phi} \rangle_E\right|^2 \right\rangle_E} \\
       \mathcal V^{(est)}_3 
	&= -\kappa_3 \frac
	{\langle r^3e^{i3\phi}\rangle_E - 3\alpha\langle r^2e^{i2\phi}\rangle_U - 6\beta\langle re^{i\phi}\rangle_C}
	{\left\langle \left|  r e^{i\phi} - \langle r e^{i\phi} \rangle_E\right|^3 \right\rangle_E}
\end{align}

In this form one can more easily see the role of the response coefficients $\kappa_n, \alpha, \beta$, and the factors that represent properties of the initial state.  
It is interesting to compare the contribution from initial transverse stress to ellpitic flow, to the  quantity known as ``momentum eccentricity'', which has been used in the past as a proxy for elliptic flow itself  at the end of the system's evolution \cite{Luzum:2009sb} and recently as an initial state estimator for  elliptic flow \cite{Giacalone:2020byk}:
\begin{align}
	\varepsilon_p &= \frac{\int d^2 x_\perp (T^{xx} - T^{yy})}{\int d^2 x_\perp (T^{xx} + T^{yy})} 
\end{align}
One can see that it almost coincides with the term $\langle 1 \rangle_C$ in the estimator proposed here.

With this Ansatz we have retained all the important properties of the estimators listed at the end of the previous section (manifest symmetries and systematic improvability), as well as one additional desired property.   While the energy density is nonzero anywhere there is matter, this is not true for momentum and stress, which can be large in principle, but can also be negligible.   Therefore we expect that if  we uniformly rescale momentum density or stress to zero, its contribution to the final anisotriopic flow should vanish.  This is in contrast to the case of energy density, where a uniform rescaling should not necessarily result in a vanishing anisotropic flow.   The proposed framework has this property, and in addition is well behaved in the case where net momentum vanishes.

We finally note that this proposal, in order to have all the generic properties that one expects for a good predictor, is quite restrictive.   Specifically, while each term in an estimator in the original framework comes with an independent unknown response coefficient, only two additional response coefficients have been added here, and a single value for each of the two coefficients should correctly describe all azimuthal harmonics, and any higher-order corrections.   This is a very restrictive condition, which can be used to test the validity of the framework.  We will return to this in Section \ref{sec:validation}.

\section{A thought experiment}\label{sec:thought}

This ansatz has all the expected and desired properties for the construction of estimators, including systematic improvements, and we will see that there is evidence from numerical simulations that the leading-order estimators predict well the results of hydrodynamic simulations.  However, it may not be obvious where it comes from.  In this section we illustrate why  Eq.~\eqref{eq:ansatz} represents a natural quantity to consider, by way of the following example.

Consider the case where at some time $\tau_0$ there are no off-diagonal elements of the energy-momentum tensor, so that the initial energy density alone is sufficient to predict the final particle distribution in a given event.  As reviewed in Section \ref{sec:energy}, we can make accurate predictions of the final flow coefficients by performing a cumulant decomposition of the initial energy density $\rho = T^{\tau\tau}(\tau_0)$, and making ratios to construct estimators.

Now imagine that we do not know exactly what is $\tau_0$, and so we want to construct estimators from the state of the system at  some time that might be slightly different, $\tau = \tau_0 + \delta\tau$.  The final particle distribution is the same, and so the estimators constructed at time $\tau$ should be close to those that we know work at time $\tau_0$.

That is, we want cumulants of the energy density at time $\tau_0$, but written in terms of quantities at time $\tau$.
%
%
%
%
%
Using local conservation of energy we can write
\begin{align}
    \partial_\tau T^{\tau\tau}(\tau) & = -\partial_i T^{\tau i}(\tau)\\
    &\simeq \frac 1 {\delta \tau} \left[T^{\tau\tau}(\tau) - T^{\tau\tau}(\tau_0) \right]
\end{align}

So our generating function can be written
\begin{align}
	\rho(\vec x_\perp)&=T^{\tau\tau}(\tau_0)\\
 &\simeq T^{\tau\tau}(\tau) - \delta\tau\partial_iT^{\tau i}(\tau)
\end{align}

Again, we know that a decomposition of this generating function gives a good estimation of the final flow.  Next, we can use the conservation of momentum to relate the momentum density to the stress tensor

\begin{align}
    \partial_\tau T^{\tau i}(\tau) &= -\partial_j T^{j i}(\tau)\\
    &\simeq \frac 1 {\delta \tau}
    \left[T^{\tau i}(\tau) - T^{\tau i}(\tau_0)\right]\\ 
    &\simeq  T^{\tau i}(\tau)/\delta\tau,
\end{align}
and so considering one higher order in $\delta \tau$ we can freely replace some or all of the quantity $-\partial_i T^{\tau i}$ with the quantity $\delta \tau\partial_i\partial_j T^{i j}$.

In this particular case, then the generating function
\begin{align}
	\rho(\vec x_\perp)=T^{\tau\tau}(\tau)-\alpha\partial_iT^{\tau i}(\tau)+\beta\partial_i\partial_jT^{ij}(\tau), 
\end{align}
produces accurate estimators for flow in the system, for any values of $\alpha$ and $\beta$ such that $\alpha + \delta\tau\beta = \delta\tau$.

In general, it will not be the case that the system is dominated by energy density at a time infinitesimally close to the considered initial time (if ever).   Thus, the response coefficients $\alpha, \beta$ will be expected to depend on the system's evolution.  Nevertheless, this thought experiment shows that the proposed generating function is a natural, robust quantity to consider, and it is not a surprise that reliable estimators can be derived from it. 

Finally, as it should be clear from the arguments used in this section, our generating function is naturally suited for hydrodynamics but it is not limited to it. The form of the generating function is obtained using the conservation of energy and momentum, which is of course more general than hydrodynamics per se. Therefore, as long as the system's dynamics is dominated by energy-momentum conservation, which should be the case for systems that satisfy our assumption 2, the generating function used in this work should be useful to determine the final-state azimuthal properties.

\section{Numerical validation}\label{sec:validation}

To test the proposed flow estimators in various collision systems, we perform state-of-the-art hydrodynamic simulations of $Pb$-$Pb$, $p$-$Pb$, and $p$-$p$ collisions, as performed at the LHC.  For initial conditions we use the IP-Glasma model \cite{Schenke:2012wb,Schenke:2012hg}, which provides a full (2D) energy-momentum tensor that can be used to initialize hydrodynamic evolution, simulated with MUSIC \cite{Schenke:2010nt}.   All hydrodynamic parameters were taken from a comprehensive Bayesian analysis \cite{Bernhard:2018hnz}.    The hydrodynamic stage is followed by the UrQMD hadronic afterburner \cite{Bass:1998ca,Bleicher:1999xi}, with oversampled events to accurately reconstruct the underlying particle distribution in every hydrodynamic event.

In each simulated event, one can compare the initial-state estimator $\mathcal V^{(est)}_n(\kappa_n, \alpha, \beta)$ proposed here with the actual flow harmonic calculated from produced hadrons $V_n$.  A natural way to quantify the event-by-event success of the estimator is to compute the Pearson correlation coefficient between the two quantities \cite{Gardim:2014tya}   
\begin{align}
\label{eq:pearson}
Q_n(\alpha,\beta) = \frac{\textrm{Re}\langle V_n\mathcal V^{(est)*}_n\rangle}{\sqrt{\langle|V_n|^2\rangle\langle|\mathcal V^{(est)}_n|^2\rangle}},
\end{align}
where in this section the angle brackets are defined as
\begin{align}
\langle\ldots\rangle = \frac{1}{N_{\textrm{events}}}\sum_{\textrm{events}}\ldots .
\end{align}
A maximal value of $Q_n = 1$ implies a perfect estimator in every event $V_n \propto\mathcal V^{(est)}_n$, while a value of 0 means they have no (linear) correlation, indicating a very poor estimator.  

The estimators \eqref{eq:estimators} depend on the usual response coefficients $\kappa_n$, associated with the response of each harmonic to the initial energy distribution, as well as two new response coeffients that represent the (relative) importance of initial momentum $\alpha$ and stress $\beta$  to the hydrodynamic response.   
However, the coefficient $\kappa_n$ is only an overall multiplicitive factor, and so it does not affect the correlation coefficient (it cancels in the numerator and denominator of Eq.~\eqref{eq:pearson}).  

To determine the best value of $\kappa_n$, we introduce 
the event-by-event error of the estimator of each harmonic
\begin{align}
\xi_n(\kappa_n,\alpha,\beta) = V_n - \mathcal V^{(est)}_n(\kappa_n, \alpha,\beta). \label{eq:Error_n}
\end{align}
We can choose $\kappa_n$ to minimize the RMS error of that respective harmonic over all the events in a centrality class.  Using the notation 
\begin{align}
 V^{(est)}_n(\kappa_n, \alpha,\beta) \equiv \kappa_n\  \varepsilon_n(\alpha,\beta),  
 \label{eq:Error}
\end{align}
%
the optimal value for a fixed $\alpha$ and $\beta$ is then
\begin{align}
\kappa_n(\alpha,\beta) = \frac{Re \langle V_n\varepsilon_n\rangle}{\langle|\varepsilon_n|^2\rangle}, \label{eq:kappa}.
\end{align}
The RMS error in this case can be written
\begin{align}
\langle|\xi_n(\alpha,\beta)|^2\rangle = \langle|V_n|^2\rangle - \kappa_n^2(\alpha,\beta) \langle|\varepsilon_n(\alpha,\beta)|^2\rangle, \label{eq:RMSError_n}
\end{align}


\begin{figure*}\centering
\includegraphics[width=\textwidth]{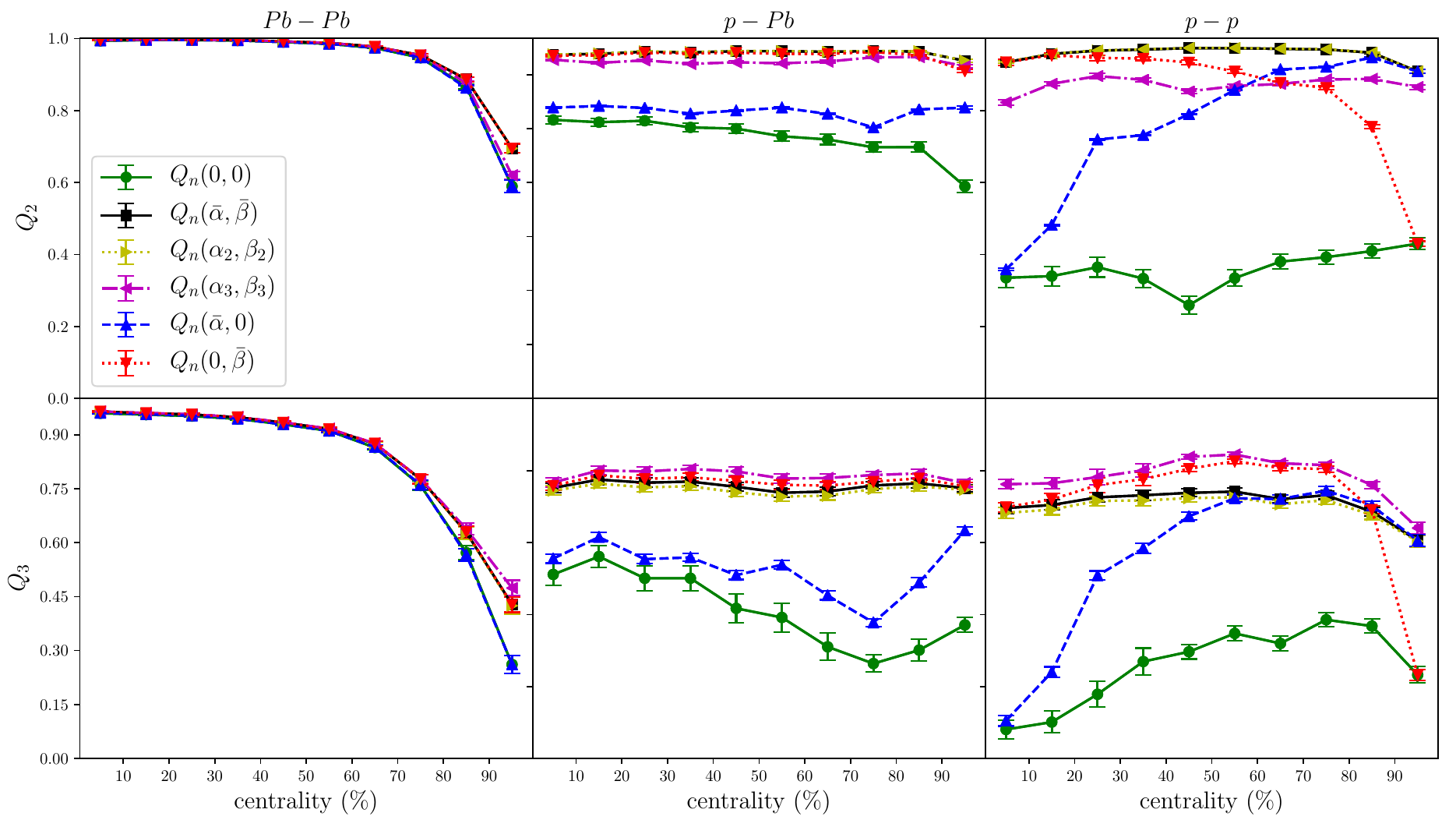}
\caption{Pearson correlation coefficient $Q_n(\alpha, \beta)$ \eqref{eq:pearson} between final flow $V_n$ and flow estimators $\mathcal V^{(est)}_n$  constructed from initial energy-momentum of hybrid hydrodynamic simulations with IP-Glasma initial conditions.  Response coefficients $(\alpha_2, \beta_2)$ and $(\alpha_3, \beta_3)$ are the values that minimize the error \eqref{eq:RMSError_n} (and maximize $Q_n$) for each harmonic, respectively, while values $(\bar \alpha, \bar \beta)$ maximize the harmonic-combined error $\eqref{eq:Error}$.  $Q_n(0,0)$ represents the usual energy-only estimator, for reference, while $Q_n(\alpha,0)$ and $Q_n(0,\beta)$ neglect only the initial stress or momentum density, respectively, to quantfy the importance of each individual contribution.  Error bars are statistical, estimated from a jackknife resampling.}
\label{Fig:FullPearson}
\end{figure*}

In Fig.~\ref{Fig:FullPearson}, we show $Q_2$ and $Q_3$ in simulated $Pb$-$Pb$, $p$-$Pb$, and $p$-$p$ systems, for various choices of response coefficients $\alpha$ and $\beta$.   The choice $\alpha = \beta = 0$ corresponds closely to the usual eccentricity, which neglects momentum degrees of freedom in the hydrodynamic initial condition.   We can see that in large systems, this eccentricity already gives an excellent estimator for the final flow coefficient, event-by-event, with $Q_n(0,0) \simeq 1$ for most centralities.  However, in smaller collision systems the usual eccentricity becomes less and less accurate as a predictor for final flow, such that $Q_n(0,0)$ is generally below 0.4 in proton-proton collisions.  This is consistent with the results of Ref.~\cite{Giacalone:2020byk}, which found that momentum degrees of freedom can become more important than spatial eccentricity in smaller systems.

We also show in Fig.~\ref{Fig:FullPearson} the Pearson correlation coefficients for the full estimators when the response coefficients are tuned to maximize each harmonic individually, which we notate as $Q_2(\alpha_2, \beta_2)$ and $Q_3(\alpha_3, \beta_3)$.   We can see that unlike the usual eccentricities the proposed estimators are excellent in all cases, achieving $Q_2>0.9$ and $Q_3 > 0.7$ even in low-multiplicity $p$-$p$ collisions.  

We note, however, that our framework does not have independent response coefficients for each harmonic, but instead share a single set of new coefficients $\alpha,\beta$.   Therefore an important non-trivial test of the proposed estimator is whether all harmonics are compatible with the same value of $\alpha$ and $\beta$. One way to test this is to compute $Q_n$ with the optimal values of $\alpha$ and $\beta$ for a different harmonic --- i.e., $Q_2(\alpha_3, \beta_3)$ and $Q_3(\alpha_2, \beta_2)$.  These are also shown in Fig.~\ref{Fig:FullPearson}.  Remarkably, the estimators still give an excellent description of the simulation results, giving very strong evidence of the validity of the proposed framework.

Finally, we define a harmonic-summed RMS error
\begin{align}
\langle|\xi(\alpha,\beta)|^2\rangle \equiv \langle|\xi_2(\alpha,\beta)|^2\rangle + \langle|\xi_3(\alpha,\beta)|^2\rangle,
\end{align}
which we can minimize by choosing optimal values $\bar \alpha$ and $\bar \beta$.
We  show $Q_2(\bar \alpha, \bar \beta)$ and $Q_3(\bar \alpha, \bar \beta)$ in Fig.~\ref{Fig:FullPearson}.  One can see that we still have an excellent estimator in all cases.

For further illustration we define a sort of likelihood function, which quantifies how good the estimator is for various values of $\alpha$ and $\beta$, using the energy-only estimator as a reference ($\alpha=\beta=0$),
\begin{align}
\mathcal L(\alpha,\beta) = \exp{\left(-\frac{\langle|\xi(\alpha,\beta)|^2\rangle}{\langle|\xi(0,0)|^2\rangle}\right)}. \label{eq:Likelihood}
\end{align}
The likelihood is maximized when the error is minimized, with an exponential decrease when the error becomes large.

\begin{figure}\centering
\includegraphics[width=\linewidth]{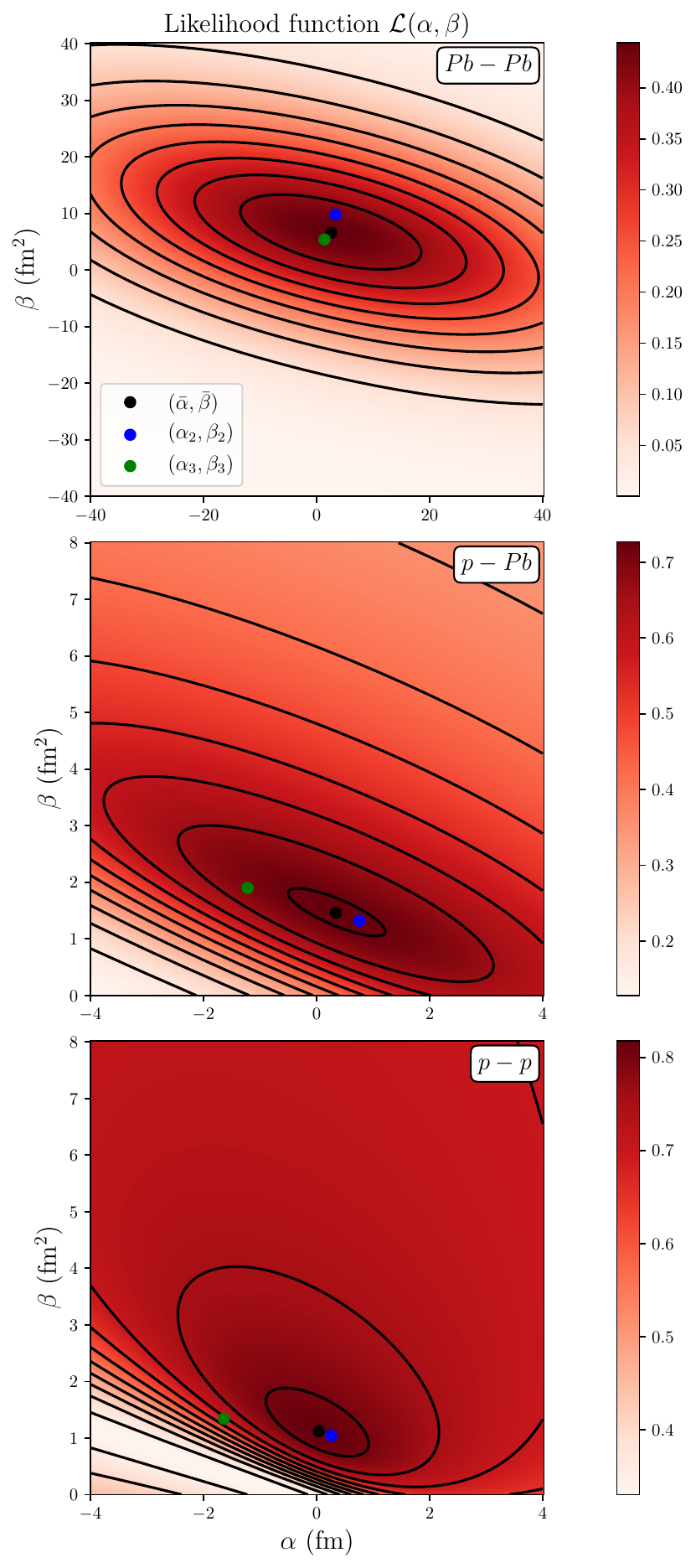}
\caption{The ``likelihood'' function, Eq.~\eqref{eq:Likelihood} as a function of response coefficients $\alpha$ and $\beta$ for the 5\% highest multiplicity events in three collision systems --- $p$-$p$, $p$-$Pb$ and $Pb$-$Pb$.  The optimal single-harmonic values ($\alpha_2$, $\beta_2$), ($\alpha_3$, $\beta_3$) have similar likelihood to the harmonic-combined optimal value ($\bar\alpha$, $\bar \beta$), giving powerful evidence for the proposed framework.
\label{Fig:Likelihoods}}
\end{figure}

In Fig.~\ref{Fig:Likelihoods}, we show a contour plot of $\mathcal L(\alpha,\beta)
$ in central collisions along with the maximal point $(\bar\alpha, \bar \beta)$ and the points that optimize each individual harmonic, $(\alpha_2, \beta_2), (\alpha_3, \beta_3)$.  

It is interesting to know which contribution is more important --- the anisotropy due to initial stress, or that of the spatial distribution of the initial momentum density (at least in this particular model of IP-Glasma + hydro).  We test this in Fig.~\ref{Fig:FullPearson} by plotting $Q_n(0, \bar \beta)$ and $Q_n(\bar \alpha, 0)$.   We see that neglecting the momentum density (setting $\alpha$=0) has a much smaller effect than the initial stress, which seems to be the dominant momentum-space hydrodynamic response except for low-multiplicity proton-proton collisions, where momentum density becomes dominant.

\section{Conclusions}

We proposed a new, systematically-improvable framework for estimating final flow coefficients $V_n$ in relativistic heavy-ion collisions from the initial energy-momentum tensor.  This extends previous work which only considered the initial energy (or entropy) density, and instead new estimators are derived that consider the system response to these additional aspects of the early-time state of the system.   Similarly to how the system response to the initial distribution of energy is contained in coefficients for each harmonic $\kappa_n$, the additional information about the system response to these other aspects of the initial condition is contained in only two additional response coefficients, $\alpha$ and $\beta$.

Using state-of-the-art hybrid hydrodynamic simulations with IP-Glasma initial conditions we found that, while the usual eccentricities become poor estimators in smaller collision systems, our new estimators remain excellent, even at lowest order in the smallest proton-proton collision systems.  Remarkably, estimators for different harmonics defined simultaneously with the same values of response coefficients $\alpha$ and $\beta$ still provide an excellent description of simulation results, providing powerful evidence of not only the utility of the estimators, but also of the correctness of the framework as a whole.

These results have several important implications.  First is the practical matter that full simulations are computationally costly.  Knowing that one is able to accurately predict final-state observables by calculating only the early-time condition of the system can potentially save a significant amount of computing time.  Beyond this, it provides insight into how different aspects of the initial condition manifest in final observables, perhaps allowing for clean separation of initial-state properties, as done with the usual eccentricity estimators in large collision systems.

Finally, the success of this framework provides some insight into the meaning of the success of hydrodynamic models in heavy-ion collisions.  While inspired by hydrodynamics, the clearly-stated postulates underlying the framework may well be more general, and it will be interesting to see whether other physical models also satisfy the postulates, and whether the various aspects of the ``collective'' paradigm of high-energy nuclear collisions are more general than hydrodynamics itself.

\section*{Acknowledgments}
We thank Jean-Yves Ollitrault and Derek Teaney for useful discussions.
This work was supported in part by FAPESP projects 2016/24029-6, 2017/05685-2 and 2018/24720-6,  by project INCT-FNA Proc.~No.~464898/2014-5, and by CAPES - Finance Code 001. J.N. is partially supported by the U.S. Department of Energy, Office of Science, Office for Nuclear Physics under Award No. DE- SC0023861.
		
\appendix

\section{List of cumulants}\label{app:cumulants}
We take the 2-D Fourier transform of the generating function  Eq.~\ref{eq:ansatz},
\begin{align}
\rho(\vec k_\perp) = \int d^2x_\perp\Bigl[T^{\tau\tau} - i\alpha k_i  T^{\tau i} - \beta k_ik_jT^{ij}\Bigr]e^{-i\vec k_\perp\cdot\vec x_\perp}.
\end{align}
and expand in a Maclaurin series around $|\vec k_\perp| = \vec 0$.  Assuming a hierarchy of importance of length scales,  we truncate at some maximum $m_{max}$
\begin{align}
\rho(\vec k_\perp) = \frac{1}{m!}\sum^{m_{max}}_{m=0}\rho_m(\phi_k)k^m.
\end{align}
In order to separate rotation modes, we decompose its Maclaurin coeficients in a Fourier series with respect to azimuthal angle ($\phi_k$)
\begin{align}
\rho(\vec k_\perp) = \frac{1}{m!}\sum^{m_{max}}_{m=0}\sum^m_{n=-m}\rho_{n,m}k^me^{-i\phi_k}.
\end{align}
The general moment can thus be expressed as
\begin{equation}
\begin{aligned}
\rho_{n,m} &= \frac{2\pi (-i)^m m!}{2^m \left(\frac{m+n}{2}\right)!\left(\frac{m-n}{2}\right)!}\int d^2x_\perp \Biggl[r^m e^{in\phi} T^{\tau\tau}\\
&-\alpha{\scriptstyle\left(\frac{m+n}{2}\right)}r^{(m-1)}e^{i(n-1)\phi}U\\
&-\alpha{\scriptstyle\left(\frac{m-n}{2}\right)}r^{(m-1)}e^{i(n+1)\phi}U^*\\
&-\beta{\scriptstyle\left(\frac{m+n}{2}\right)\left(\frac{m+n}{2}-1\right)} r^{(m-2)}e^{i(n-2)\phi}C\\
&-\beta{\scriptstyle\left(\frac{m+n}{2}\right)\left(\frac{m-n}{2}-1\right)} r^{(m-2)}e^{i(n+2)\phi}C^*\\
&-\beta{\scriptstyle\left(\frac{m+n}{2}\right)\left( m-n\right)}r^{(m-2)}e^{in\phi}T\Biggr],
\end{aligned}
\end{equation}
where, $T$, is the following trace written as
\begin{align}
T = T^{xx} + T^{yy}.
\end{align}
Note, however, that the trace does not contribute to the leading estimators, where $n=m$.

To obtain translation-invariant cumulants, we define a logarithmic function of $\rho(\vec x_\perp)$,
\begin{align}
W(\vec k_\perp) \equiv \ln\Bigl[\rho(\vec k_\perp)\Bigr],
\end{align}
and, similarly, we expand it in Maclaurin and Fourier series
\begin{align}
W(\vec k_\perp) = \sum^{m_{max}}_{m=0}\sum^m_{n=-m} W_{n,m}k^me^{-i\phi_k}.
\end{align}
Then, some of the lowest cumulants can be written as
\begin{equation}
W_{1,1} = \frac{(-i)}{2}\Bigr[\langle re^{i\phi}\rangle_E - \alpha\langle1\rangle_U\Bigr],
\end{equation}
\begin{equation}
\begin{aligned}
W_{2,2} = \frac{(-i)^2}{4}\Biggr[&\langle r^2e^{i2\phi}\rangle_E - 2\alpha\langle re^{i\phi}\rangle_U - 2\beta\langle 1\rangle_C \\   &-(\langle re^{i\phi}\rangle_E - \alpha\langle 1\rangle_U)^2\Biggr],\\
\end{aligned}
\end{equation}
\begin{equation}
\begin{aligned}
W_{3,3} = \frac{(-i)^3}{8}\Biggr[&\langle r^3e^{i3\phi}\rangle_E - 3\alpha\langle r^2e^{i2\phi}\rangle_U - 6\beta\langle re^{i\phi}\rangle_C\\
&-\Bigr(\langle re^{i\phi}\rangle_E - \alpha\langle 1\rangle_U\Bigr) \\
&\cdot\Bigr(3(\langle r^2e^{i2\phi}\rangle_E - 2\alpha\langle re^{i\phi}\rangle_U - 2\beta\langle 1\rangle_C) \\ &-2(\langle re^{i\phi}\rangle_E - \alpha\langle 1\rangle_U)^2\Bigr)\Biggr].
\end{aligned}
\end{equation}

Cumulants of higher order have a rapidly-increasing number of terms.  We include a few here, simplified by writing in terms of moments $\rho_{n,m}$.
\begin{equation}
\begin{aligned}
W_{1,3} = \frac{3(-i)^3}{8} \Biggl[& \langle r^3e^{i\phi} \rangle_E - 2\alpha\langle re^{i\phi}\rangle_U
- 4\beta\langle re^{-i\phi}\rangle_C - 4\beta\langle re^{i\phi}\rangle_T \\
&- \rho_{2,2}\rho_{-1,1} - 2\rho_{0,2}\rho_{1,1} + 2\rho_{1,1}^2\rho_{-1,1},\Biggr]
\end{aligned}
\end{equation}

\begin{equation}
\begin{aligned}
W_{4,4} = \frac{(-i)^4}{16} \Biggl[& \langle r^4e^{i4\phi} \rangle_E - 4\alpha\langle r^3e^{i3\phi}\rangle_U
- 12\beta\langle r^2e^{-i2\phi}\rangle_C\\ 
&-4\rho_{3,3}\rho_{1,1} - 3\rho_{2,2}^2 + 12\rho_{2,2}\rho_{1,1}^2 - 6\rho_{1,1}^4,\Biggr]
\end{aligned}
\end{equation}

\begin{equation}
\begin{aligned}
W_{5,5} = \frac{(-i)^5}{32} \Biggl[& \langle r^5e^{i5\phi} \rangle_E - 5\alpha\langle r^4e^{i4\phi}\rangle_U
- 20\beta\langle r^3e^{-i3\phi}\rangle_C\\ 
&+ 20\rho_{3,3}\rho_{2,2}^2 - 60\rho_{2,2}\rho_{1,1}^3 + 5\Bigr(6\rho_{2,2}^2-\rho_{4,4}\Bigr)\rho_{1,1}\\
&- 10\rho_{2,2}\rho_{3,3} + 24\rho_{1,1}^5\Biggr]
\end{aligned}
\end{equation}


\end{document}